\documentclass[runningheads, anonymous]{llncs}
\usepackage[T1]{fontenc}

\AtBeginDocument{%
  }
\usepackage{url}

\usepackage{breakurl}
\usepackage{hyperref}
\usepackage{subcaption}
\usepackage{textcomp}

\usepackage{comment}
\usepackage{xcolor}
\usepackage{glossaries}
\usepackage{svg}
\usepackage{algorithm}
\usepackage{multirow}

\usepackage{graphicx}
\usepackage{booktabs,arydshln}
\usepackage{pdflscape}
\usepackage{subcaption}
\usepackage{ragged2e}
\usepackage{algorithm}
\usepackage{algpseudocode}
\usepackage{cleveref}
\usepackage{siunitx}
\usepackage{tikz}
\usetikzlibrary{arrows.meta,decorations.pathreplacing,positioning}
\usepackage{enumitem}
\usepackage{adjustbox}
\usepackage{pdflscape}
\usepackage{subcaption}
\usepackage{balance}
\usepackage{cite}
\usepackage{todonotes}
\usepackage{multirow}
\makeatletter
\def\adl@drawiv#1#2#3{%
        \hskip.5\tabcolsep
        \xleaders#3{#2.5\@tempdimb #1{1}#2.5\@tempdimb}%
                #2\z@ plus1fil minus1fil\relax
        \hskip.5\tabcolsep}
\newcommand{\cdashlinelr}[1]{%
  \noalign{\vskip\aboverulesep
           \global\let\@dashdrawstore\adl@draw
           \global\let\adl@draw\adl@drawiv}
  \cdashline{#1}
  \noalign{\global\let\adl@draw\@dashdrawstore
           \vskip\belowrulesep}}
\makeatother
\glsdisablehyper

\title{Arbitrage with bounded liquidity}
\author{Christoph Schlegel and Quintus Kilbourn}
\institute{Flashbots}
\date{}
\begin{document}
%
      
%
%
%
%
%
\maketitle              
\begin{abstract}
We derive the arbitrage gains or, equivalently, Loss Versus Rebalancing (LVR) for arbitrage between \textit{two imperfectly liquid} markets, extending prior work that assumes the existence of an infinitely liquid reference market. Our result highlights that the LVR depends on the relative liquidity and relative trading volume of the two markets between which arbitrage gains are extracted. Our model assumes that trading costs on at least one of the markets is quadratic. This assumption holds well in practice, with the exception of highly liquid major pairs on centralized exchanges, for which we discuss extensions to other cost functions.
\end{abstract}

\section{Introduction}
Contrary to popular belief, arbitrageurs cannot trade against infinite liquidity at Binance mid price. Arbitrageurs face costs and ignoring this cost leads to systematic overestimation of their profits.
The first sentence is, of course, a straw man: Loss versus Rebalancing (LVR) is a measure of the value leakage of passive liquidity providers (LPs) on DEXes (decentralized exchanges) to other actors. The profile of these actors and the value distribution between them has not been the focus of existing theory. Extending theory to address these questions serves several important functions. Firstly, knowing how value is distributed among passive liquidity providers and active market makers across different venues, arbitrageurs and underlying platforms gives a more holistic picture upon which to found further market structure research. Secondly, there are equilibrium effects: in the long run the liquidity distribution across exchanges should reflect the relative gains to liquidity provisioning across these venues. If LPs on one venue lose relative to LPs on another venue, e.g. if one group internalizes the LVR that the other leaks, this should be reflected in the market eventually. Third, the standard analysis usually mixes two effects for the reference market, which are in reality confounded but separate: there is different liquidity in the two markets \emph{and} liquidity provisioning on the reference market is active, while DEX LPs are largely passive. It would be instructive to disentangle the effect of these two assumptions.  

In the original definition, given in \cite{milionis2022}, losses of LPs in one market are studied under the assumptions of the existence of another perfectly liquid ideal market, reflecting an enormous assumed difference between DEXes and centralized exchanges (CEXes). Our work relaxes this assumed distinction and instead studies dynamics arising out of the economic properties of trading venues - liquidity levels and information arrival processes. Our core assumption is that trading cost (slippage) on at least one of the markets is quadratic. As this assumption holds broadly in practice, our theory applies to a large class of settings, including those involving multiple AMMs \cite{oz2025crosschainarbitragefrontiermev}. The notable exception is highly liquid pairs on (CEX) limit order books, as explained in the next section.

Our model predicts that LVR is determined by the relative liquidity provisioned \emph{at the margin} around the equilibrium exchange rate and the relative trading volume. One implication is that, all else being equal, more LVR is leaked, the more unbalanced liquidity and volume is between the markets.  Another implication is that active liquidity providers can anticipate changes in equilibrium exchange rates to determine how much liquidity to provide at the margin. This would reduce their LVR and increase LVR for passive LPs.
\subsubsection*{Background}
The study of LVR was motivated by the rise of automated market maker (AMM) smart contract exchanges on public blockchains - i.e. DEXes~\cite{univ2, univ4Overview,dexSoK,canidio2025arbitrageursprofitslvrsandwich}. DEXes aim to facilitate trade without reliance on any central authority. This is achieved by making use of a distributed blockchain network. As a consequence, DEXes inherent some properties of the underlying infrastructure such as the execution of trades at discrete time intervals (e.g. every 12s on Ethereum) and high fees \cite{fan2024strategicliquidityprovisionuniswap}, which, in turn, shape the economic dynamics of these markets. In particular, liquidity provisioning is typically \textit{passive} in comparison with traditional limit order book CEXes. AMM logic programmatically adjusts quoted prices as a function of executed trades (recent designs implement more complex mechanisms). This leads to LPs on DEXes quoting outdated prices, motivating the study of their profitability through measures such as LVR. 

Our work can be interpreted as part of a larger effort to connect theory and practice. Since the first rigorous study of arbitrage on decentralized exchanges in \cite{daian2019flashboys20frontrunning}, there has been an ongoing effort to model the economic behaviour of both liquidity providers and arbitrageurs in these markets. After LVR was introduced in \cite{milionis2022} (improving upon previous methods for measuring loss \cite{angeris2020doestailwagdog}), the authors published a refinement accounting for fees charged by liquidity providers \cite{milionis2024}. Subsequent work modeled different facets of decentralized exchanges such as infrastructure fees (``gas'') \cite{he2025arbitragedecentralizedexchanges}, and update frequency (block times) \cite{nezlobin2025}. 
 Empirical works have estimated arbitrage gains and LVR in practice \cite{fritsch2024measuringarbitragelossesprofitability,wu2025measuring} assuming that arbitrageurs can hedge their trades at almost zero cost on an external market.

Most similarly to our work, \cite{willetts2024rebalancingversusrebalancingimprovingfidelitylossversusrebalancing} proposes a refinement of LVR, RVR, which accounts for trading costs on CEXes in the reference rebalancing strategy. This resembles our model which also accounts for costs on both markets. In contrast to our work, the analysis in \cite{willetts2024rebalancingversusrebalancingimprovingfidelitylossversusrebalancing} is based on simulations and empirical estimations instead of deriving theoretical closed-form expressions for LVR with trading cost on both venues, as we do in our analysis.

 Our model also follows on the empirical observations of arbitrage across decentralized venues with limited liquidity in \cite{oz2025crosschainarbitragefrontiermev}. To the best of our knowledge, this is the first analytical work that considers LVR where all venues have trading costs.

We close this paper with potential extensions of the model and open research questions.

\section{Model and Results}\label{model}
There are two venues where a pair of tokens $A$ and $B$ is traded. Token $B$ is the numéraire. Trading on the first venue follows the setting of the LVR paper, where $x^*(Q)$ denotes equilibrium reserves of the risky-asset at exchange rate $Q$.

In contrast to the standard model, we assume that the second venue also has bounded liquidity $\tilde{x}^*(Q)$ and that trading on the second venue has a quadratic trading cost that depends on the liquidity as well as the price. More specifically, the trading cost for $\Delta x$ amount of token $A$ is
$$C(Q,\Delta x)=\frac{(\Delta x)^2}{2|\tilde{x}^{*\prime}(Q)|},$$
i.e. selling $\Delta x$ amount of token $A$ yields $Q\Delta x-C(Q,\Delta x)$ amount of token $B$ (rather than $Q\Delta x$ if there wasn't a trading cost).\footnote{In the most straightforward interpretation, our model of trading cost assumes that the arbitrageur hedges immediately. This is not an exact model of strategies seen in practice, since accumulating inventory and waiting for counterbalancing trades reduces total hedging costs. Our model and the existing DeFi literature does not consider this optimal hedging problem. 
While hedging cost in reality is multi-factorial and does not only depend on instantaneous slippage, our model can also apply to more sophisticated strategies in the presence of quadratic hedging cost.} 

We assume that the equilibrium market exchange rate after the market has been arbitraged to efficiency follows a GBM
$$\frac{d Q_t}{Q_t}\equiv\sigma dB_t$$
for a Brownian motion $B_t$. This is the same price dynamics assumption as in the standard model. In contrast to previous works, our model dictates that \textit{both} markets can be out of equilibrium whenever arbitrage is possible, and arbitrageurs bring the markets to the correct exchange rate, incorporating all available information.
\begin{remark}
We can micro-found this specification of prices as follows: Suppose that noise traders' trades appear with some probabilities on either of the two exchanges. For each price multiple $u$, we can always find a quantity $q$ such that, \textit{independently of the venue} on which the noise trader buys $q$ of the risky asset, an arbitrageur maximizing his numéraire holding will bring both markets to the post trade price $p*u$ where $p$ is the price prior to the noise trade. For example, for a constant product market maker (CPMM), $q=(1-1/\sqrt{u})(x(p)+\tilde{x}(p))/p.$

Consider a binomial tree type model, where the noise trader buys $q$ of the risky asset (resp. sells $\tilde{q}$) such that the price post-arbitrage moves by a factor $u=e^{\sigma\sqrt{\Delta t}}$ (resp. $\tilde{u}=1/u$) with probabilities $1/(e^{\sigma\sqrt{\Delta t}}+1)$ (resp. $e^{\sigma\sqrt{\Delta t}}/(e^{\sigma\sqrt{\Delta t}}+1)$). Applying Donsker's theorem to log prices, we get a GBM price process in the limit.

In other words, we justify our price process by assuming that the non-arbitrage trades (which determine price movements) are sized according to a particular function of the price at the time of their arrival, and that risk-free arbitrage happens whenever possible. 
\end{remark}


\subsection*{Why quadratic trading cost?}
As in previous literature, we abstract away fees, spreads and other costs to focus on slippage.
We justify the quadratic trading cost formula for slippage as follows: taking a linear approximation,
$$\tilde{x}(\tilde{Q})\approx |\tilde{x}'(Q)|(\tilde{Q}-Q),$$ so that a trader trading $\Delta x$ pays approximately
$$Q\Delta x+\int_0^{\Delta x}\tfrac{1}{|\tilde{x}'(Q)|}xdx=Q\Delta x+\frac{1}{2|\tilde{x}'(Q)|}(\Delta x)^2,$$
the cost of trading is $$C(Q,\Delta x)\approx\frac{1}{2|\tilde{x}'(Q)|}(\Delta x)^2.$$

\subsubsection*{Costs On AMMs}
 Suppose market $2$ is a CPMM. CPMMs are a particular kind of AMM that is based on the constraint that the reserves of the risky asset $\tilde{x}(Q)$ and of the numéraire $\tilde{y}(Q)$ at a quoted marginal exchange rate $Q$ satisfy $\tilde{x}(Q)*\tilde{y}(Q)=\tilde{K}^2$ where $\tilde{K}>0$ is a constant. Then the trading cost (of selling $\Delta x$ of the risky asset) is given by
$$C=Q \Delta x-\Delta y$$
where  $$(\tilde{x}^*(Q)+\Delta x)(\tilde{y}(Q)-\Delta y)=\tilde{K}^2=\tilde{x}^*(Q)\tilde{y}(Q)\Rightarrow\Delta y=\frac{\tilde{y}(Q)\Delta x}{\tilde{x}(Q)+\Delta x}.$$
Thus
\begin{align*}C&=Q \Delta x-\frac{\tilde{y}(Q)\Delta x}{\tilde{x}(Q)+\Delta x}=\frac{Q(\Delta x)^2}{\tilde{x}(Q)+\Delta x}\approx \frac{Q(\Delta x)^2}{\tilde{x}(Q)}.\end{align*}
The last approximation is good, as long as the trade size $\Delta x$ is small relative to the pool reserves $\tilde{x}^*(Q)$. 
For the purpose of our main theorems, we have in particular that the trading cost of arbitrageurs over time is the same on a CPMM and a trading venue with quadratic cost as long as the price process is a GBM:
\begin{proposition}
For a CPMM cost function $C(\Delta x):=\frac{Q(\Delta x)^2}{x(Q)+\Delta x}$ and quadratic cost $\tilde{C}(\Delta x):=\frac{Q(\Delta x)^2}{x(Q)}$ we have
$$\lim_{N\to\infty}\sum_{i=1}^NQ_{t_{i-1}}\frac{(x_{i_{i-1}}-x_{i_{i}})^2}{\tilde{x}_{t_{i-1}}}=\lim_{N\to\infty}\sum_{i=1}^NQ_{t_{i-1}}\frac{(x_{i_{i-1}}-x_{i_{i}})^2}{\tilde{x}_{t_{i-1}}+(x_{i_{i-1}}-x_{i_{i}})}$$

\end{proposition}
\begin{proof}
Taylor expanding the cost function of a CPMM:
$$\frac{Q(\Delta x)^2}{x(Q)+\Delta x}=0+0+\frac{Q}{x(Q)}(\Delta x)^2+\frac{Q(\Delta x)^3}{x(Q)^2}+\ldots.$$
 Note that for each $k>0$ there is $\bar{C}>0$ such that the $k$-the derivative of the CPMM cost functions is a.s. bounded by it, i.e. $C^{(k)}(Q_t,0)<\bar{C}$ a.s. for all $t\leq T$. Note moreover that $x(Q_t)$ is a.s. path-wise continuous so that $\sup_{i=1,\ldots,\infty}|x(Q_{t_{i-1}})-x(Q_{t_{i}})|=0$. Thus, for the cubic and higher order terms, using that $x(Q)$ has finite quadratic variation $[x(Q)]_{T}$ we have
\begin{align*}\lim_{N\to\infty}\tfrac{1}{k!}\sum_{i=1}^NC^{(k)}(Q_{t_{i-1}},0)(x_{i_{i-1}}-x_{i_{i}})^k\leq \lim_{N\to\infty}\frac{\bar{C}}{k!}\sum_{i=1}^N(x_{t_{i-1}}-x_{t_{i}})^{k}\\\leq\tfrac{\bar{C}}{k!}\sup (x_{t_{i-1}}-x_{t_i})^{k-2}[x(Q)]_{T}=0.\end{align*}
Thus for the limit
$\tilde{C}(\Delta x):=\frac{Q(\Delta x)^2}{x(Q)}$ we have
$$\lim_{N\to\infty}\sum_{i=1}^NQ_{t_{i-1}}\frac{(x_{i_{i-1}}-x_{i_{i}})^2}{\tilde{x}_{t_{i-1}}+(x_{i_{i-1}}-x_{i_{i}})}=\lim_{N\to\infty}\sum_{i=1}^NQ_{t_{i-1}}\frac{(x_{i_{i-1}}-x_{i_{i}})^2}{\tilde{x}_{t_{i-1}}}$$
\end{proof}

Practically speaking, Uniswap V2 and V3 are canonical AMM designs that are representative of many DEX deployments. Uniswap V2 is a CPMM so the argument above goes through, and local trading costs on Uniswap V3 are also known to be approximately quadratic \cite{UniV3, axiomsAMM}.

\subsubsection*{Costs on CLOBs}
Practically, we can infer the slope $|\tilde{x}'(Q)|$ on order books from data, by running a regression around market mid-price. 
Linear marginal cost is a good approximation for small trade sizes in a limit order book in some situations, and is not uncommon in finance literature (e.g. \cite{almgren_chriss_2001}). Empirically, the approximation tends to work better for smaller market cap pairs, partially for apparently idiosyncratic reasons: minimum tick sizes tend to be more granular for these pairs, whereas for large cap pairs, such as ETH/USDT and BTC/USDT on Binance, tick sizes are relatively large (1 cent). This leads to the effect that a lot of liquidity is concentrated at the top of book and liquidity is sparser elsewhere (for adverse selection effects). For these token pairs, a more reasonable model of cost is to assume constant rather than linear marginal cost up to a certain trade size and non-constant cost above that. We briefly discuss this setting in Section~\ref{Cost}.

\subsection*{Defining Loss versus Rebalancing}
The Loss versus Rebalancing is, as in the standard framework, the difference in value between the portfolio value of the LPs and a counter-factual rebalancing strategy where we would re-balance the portfolio at equilibrium exchange rates:
$$LVR_T:=R_T-V_T.$$
The rebalancing strategy is now, however, subject to trading costs: More precisely, we have
\begin{align*}R_T&=V_0+\int_0^Tx^*(Q_s)dQ_s-\int_0^T \frac{1}{2|\tilde{x}'(Q_s)|}d[x^*(Q)]_s\\&=V_0+\int_0^Tx^*(Q_s)dQ_s-\int_0^T \frac{\sigma^2Q_s^2}{2|\tilde{x}'(Q_s)|}(x^{*\prime}(Q_s))^2ds\end{align*}
where $[x^*(Q)]$ denotes the quadratic variation of $x^*(Q)$ and its expression follows from
It\^{o}'s Lemma.

\subsection*{Calculating Loss versus Rebalancing}
We next calculate LVR for our setting. The formula for the LVR can be adjusted by a term that depends on the relative liquidity at the margin. As in the original setting, LVR is equal to the arbitrage in a competitive market with many arbitrageurs. For the case that the second market is the leading market where price discovery happens, we get the following:
\begin{theorem}
For the bounded liquidity case, LVR and arbitrage gains ARB over the interval $[0,T]$ take the form,
$$\text{LVR}_T=\text{ARB}_T=\int_0^T\ell(\sigma,Q_t)dt$$
where
$$\ell(\sigma,Q):=\tfrac{\sigma^2Q^2}{2}(1-\tfrac{|x^{*\prime}(Q)|}{|\tilde{x}^{*\prime}(Q)|})|x^{*\prime}(Q)|.$$
\end{theorem}
\begin{proof}

With $N$ arbitrageurs the arbitrage gains on the time interval $[0,T]$ is given by 
\begin{align*}
ARB_T^{(N)}=\sum_{i=1}^NQ_{t_{i}}(x^*_{t_{i-1}}-x^*_{t_i})-(y^*_{t_{i}}-y^*_{t_{i-1}})-\frac{1}{2\tilde{x}^{\prime}_{t_{i}}}\left(x^*_{t_{i-1}}-x^*_{t_{i}}\right)^2, 
\end{align*}
where the expression is the same as in \cite{milionis2022} except for the last term which is the quadratic trading cost. When going to the limit, the first two terms become (as in the calculation in \cite{milionis2022}),
\begin{align*}&\lim_{N\to\infty}\sum_{i=1}^NQ_{t_{i}}(x^*_{t_{i-1}}-x^*_{t_i})-(y^*_{t_i}-y^*_{t_{i-1}})= V_0+\int_0^T|x^*(Q_t)|dQ_t-V_T\end{align*}
For the third term we get the following limit:
\begin{align*}\lim_{N\to\infty}\sum_{i=1}^N \frac{1}{2\tilde{x}^{\prime}_{t_{i}}}\left(x^*_{t_{i-1}}-x^*_{t_i}\right)^2=\int_0^T \frac{1}{2\tilde{x}^{*\prime}(Q_t)}d[x^*(Q)]_t.\end{align*}
It follows that 
$$ARB_T=V_0+\int_0^T|x^*(Q_t)|dQ_t-\int_0^T \tfrac{1}{2\tilde{x}^{*\prime}(Q_t)}d[x^*(Q)]_t- V_T=R_T-V_T=LVR_T$$
By It\^{o}'s formula, we obtain
$$d[x^*(Q)]_t={\sigma^2 Q_t^2}(x^{*\prime}(Q_t))^2dt.$$
Thus the cost term is
$$\int_0^T\frac{\sigma^2Q_t^2}{2\tilde{x}^{*\prime}(Q_t)}(x^{*\prime}(Q_t))^2dt$$
and we obtain the adjusted formula for the arbitrage gain:
$$ARB_T=V_0-V_T+\int_0^T|x^*(Q_t)|dQ_t-\int_0^T\frac{\sigma^2Q_t^2}{2\tilde{x}^{*\prime}(Q_t)}(x^{*\prime}(Q_t))^2dt.$$
By It\^{o}'s lemma and using the Envelope theorem:
$$V_T-V_0=\int_0^TV'(Q_t)dQ_t+\int_0^T\tfrac{1}{2}V''(Q_t)\sigma^2Q_t^2dt=\int_0^Tx^*(Q_t)dQ_t+\int_0^T\tfrac{1}{2}x^{*\prime}(Q_t)\sigma^2Q_t^2dt$$
\end{proof}
\subsubsection*{Price discovery on either market}
A generalization of the previous results is to the case where price discovery does not exclusively happen on one venue.
The GBM price movements are modeled as arising from the process by which trades (order flow) arrive. As argued in Remark~1, the price movement depends on trade size and direction, but not the venues where the trades execute. In the following, we model the order flow distribution by the condition that a fraction $\pi$ of all orders go to the first market and a fraction $1-\pi$ to the second market. More precisely, in the multiplicative random walk as sketched in Remark~1, we assume that noise traders appear independently over time with probability $\pi$  on market $1$ and with probability $1-\pi$ on market $2$, so that in the limit almost surely a fraction $\pi$ of noise traders appears on market $1$ and a fraction  $1-\pi$ on market $2$. This distribution can be interpreted as arising from relative block times between two CEXes or chosen to be reflective of differing liquidity levels.
\begin{theorem}
In case the two markets use CPMMs with order flow distribution $(\pi,1-\pi)$, we have 
$$\text{ARB}_T=\int_0^T\ell(\sigma,Q_t)dt\quad\text{a.s.}$$
with
$$\ell(\sigma,Q)=\tfrac{\sigma^2Q^2}{2}|\tfrac{1}{|x'(Q)|}-\tfrac{1}{|\tilde{x}'(Q)|}|\left(\pi|\tilde{x}'(Q)|^2 +(1-\pi)|x'(Q)|^2\right).$$
\end{theorem}
\begin{proof}
As discussed before, we can approximate the total revenue of selling $\Delta x$ on market $1$ with initial marginal exchange rate $Q^{old}$ by
$$Q^{old}\Delta x-\frac{1}{2|x'(Q^{old})|}\Delta x^2.$$
The marginal exchange rate post trade $Q^{new}$ is given by the derivative of the trading revenue, i.e.
$$Q^{new}=Q^{Old}-\frac{1}{|x'(Q^{old})|}\Delta x.$$
Analogously, for the second market, we can approximate the total cost of buying $\Delta x$ on market $2$ with initial marginal exchange rate $Q^{old}$ by
$$Q^{new}=Q^{old}+\frac{1}{|\tilde{x}'(Q^{old})|}\Delta x.$$
Following our model of noise trading in Remark~1, the profit of the arbitrageur for the arbitrage trade after the noise trader has moved is
\begin{align*}&Q^{new}\Delta x+\frac{1}{|x'(Q^{old})|}\Delta x^2-\frac{1}{2|x'(Q^{old})|}\Delta x^2-(Q^{new}\Delta x+\frac{1}{|\tilde{x}'(\tilde{Q}^{old})|}\Delta x^2-\frac{1}{2|x'(\tilde{Q}^{old})|}\Delta x^2)\\=&\frac{1}{2}\Delta x^2\left(\frac{1}{|x'(Q^{old})|}-\frac{1}{|\tilde{x}'(\tilde{Q}^{old})|}\right)\end{align*}
Suppose noise trades appear at times $t_1,\ldots,t_n$ on the time interval $[0,T]$ and arbitrageurs back run them immediately afterwards. Let $m_i$ denote an indicator variable that is $1$ if the noise trader at time $t_i$ trades on market $1$ and $2$ if the noise trader trades on market $2$.
The total arbitrage profit is given by
\begin{align*}\sum_{i=1}^n {1}_{\{m_i=1\}}(\tilde{x}(Q_{t_{i-1}})-\tilde{x}(Q_{t_i}))^2|\tfrac{1}{|x'(\hat{Q}_{t_{i}})|}-\tfrac{1}{|\tilde{x}'({Q}_{t_{i-1}})|}|\\+\sum_{i=1}^n1_{\{m_i=2\}}(x(Q_{t_{i-1}})-{x}(Q_{t_i}))^2|\tfrac{1}{|x'(Q_{t_{i-1}})|}-\tfrac{1}{|\tilde{x}'(\hat{Q}_{t_{i}})|}|,\end{align*}
where $\hat{Q}$ denote the price (on the market touched by the noise trader) after the noise trade  but prior to the arbitrage trade.
We now consider the limit of the profit when going to the functional central limit as described in Remark~1. For noise traders arriving continuously, we have by the LLN  $\sum_{i=1}^\infty {1}_{\{m_i=1\}}=\pi$ and $\sum_{i=1}^\infty {1}_{\{m_i=2\}}=1-\pi$ almost surely. Thus the arbitrage profit converges to 
$$\int_0^T \tfrac{1}{2}\pi|\tfrac{1}{|x'(Q_t)|}-\tfrac{1}{|\tilde{x}'(Q_{t})|}| d[\tilde{x}(Q)]_t+\int_0^T \tfrac{1}{2}(1-\pi)|\tfrac{1}{|x'(Q_{t})|}-\tfrac{1}{|\tilde{x}'(Q_{t})|}| d[x(Q)]_t.$$
By It\^{o}'s formula, we obtain
$$d[x(Q)]_t={\sigma^2 Q_t^2}(x^{\prime}(Q_t))^2dt\quad\text{and}\quad d[\tilde{x}(Q)]_t={\sigma^2 Q_t^2}(\tilde{x}^{\prime}(Q_t))^2dt.$$
Thus, 
$$ARB_T=\int_0^T \tfrac{\sigma^2Q_t^2}{2}|\tfrac{1}{|x'(Q_t)|}-\tfrac{1}{|\tilde{x}'(Q_{t})|}|\left(\pi|\tilde{x}'(Q_{t})|^2 +(1-\pi)|x'(Q_{t})|^2\right)dt,$$
where all previous equations hold almost surely.
Finally, we note that using the approximation $\frac{Q(\Delta x)^2}{x(Q)+\Delta x}\approx \frac{Q(\Delta x)^2}{x(Q)}$ for the CPMM cost does not change the formulae in the limit, as cubic and higher order terms of the Taylor expansion do not change the evaluation of the Ito integrals. Thus, the arbitrage gains formula does not only hold approximately but exactly for two CPMMs.\qed
\end{proof}
\subsection*{Interpretation and Discussion}
Let us first look at the DEX-DEX case from Theorem~2. Assuming that both DEXes are CPMMs with $\pi=0$ and normalizing by liquidity provided we get:
$$\frac{\ell}{V}=\tfrac{\sigma^2}{8}(2\pi-1)(1-1/r),\text{ with }r:=\tfrac{\tilde{x}}{x}.$$
The returns to liquidity provisioning are higher the more balanced liquidity and trading volume is between the two markets. 

Theorem~1 tells us, moreover, that there are gains to be made from active market making. This is, of course, not surprising at all qualitatively, but we get a quantitative estimation of this effect. What determines the relative LVR in the two exchanges is the relative marginal liquidity provided in the two exchanges at equilibrium price.
If LPs actively manage liquidity by trying to predict the evolution of equilibrium exchange rates and react to it by adjusting supply, they are naturally better off compared to an LP that commits to a supply curve $x^*$ and that independently of how much capital they can deploy. 
\section{Extensions \& Open Research Questions}\label{Cost}
\subsubsection*{Other Cost Functions}
The previous results are dependent on the shape of the cost function. As discussed previously, for high volume tickers on Binance, linear marginal cost is a bad approximation and we should work with the following model instead:
$$C'(\Delta x)=\begin{cases}
c,\quad  &\Delta x\leq \alpha ,\\
\tfrac{\Delta x-\alpha}{|x'(Q)|}+c,\quad&\Delta x> \alpha,
\end{cases}$$
for constants $\alpha>0$ and $c>0$.
At this point our model assumption of a diffusion process becomes crucial and we would obtain quite different results otherwise. 
For the GBM model, almost surely all arbitrage trades are ``small" i.e. they are below the $\alpha$ threshold so that we can work with a constant marginal cost assumption.
If we would literally follow the logic of the LVR calculation, instead of quadratic variation, we would need to look at the absolute variation for the cost,
$\lim_{N\to\infty}\sum_{i=1}^N c\left|x^*_{t_{i-1}}-x^*_{t_i}\right|$,
which diverges. Hence, the result does not extend.  Of course, arbitraging not being worthwhile directly contradicts reality and is partly a consequence of the assumption that hedging is done immediately. 

DEX trading costs can also vary. Decentralized infrastructure and competition for arbitrage opportunities can impose additional costs on arbitrageurs in the form of fees or auction bids for priority (\cite{daian2019flashboys20frontrunning, solmaz2025optimisticmevethereumlayer,he2025arbitragedecentralizedexchanges}). Fortunately, other constant function market maker designs have already been studied in related settings (\cite{CFMM,goyal2023findingrightcurveoptimal}).
\subsubsection*{Other Price Processes}
One difference between our model and existing LVR literature is the way in which the reference price is derived. Other notions of reference price would expand the settings in which LVR can be calculated. For example, calculating adverse selection loss for assets that are only traded on one DEX could perhaps be modeled by arbitrageurs that only have noisy access to a reference price.

\subsubsection*{Finite Slot Times}
The LVR framework has been extended to the case of fees, where arbitrageurs are restricted to trade on chain at discrete (possibly random) slot times~\cite{milionis2024,nezlobin2025}. It would be relatively straightforward to extend their analysis by introducing a slippage cost while maintaining the no-spread assumption for the more liquid market. However, this is not a reasonable model. Ideally, we would have a model that would have fees/spreads in both markets. Such extensions seem non-trivial to analyze as there would no longer be a single equilibrium exchange rate. Rather, there is a corridor of exchange rates where there is no profitable arbitrage. Coming up with a tractable model is an interesting open question. 
\subsubsection*{Empirical Research}
Realistic estimates of arbitrageur as well as liquidity provider (resp. market maker) P\&L and cost of liquidity are arguably largely missing from the DeFi literature.  Existing literature such as~\cite{wu2025measuring} seem to systematically overestimate the profits of arbitrageurs. Thus, it would be interesting to construct a reasonable multi-factor model of cost for these actors and to estimate it from public data.

%
%
%
%
%
%
\bibliographystyle{splncs04}
\bibliography{name}
\appendix

\end{document}